\documentclass{IEEEcsmag}

\usepackage[colorlinks,urlcolor=blue,linkcolor=blue,citecolor=blue]{hyperref}
\expandafter\def\expandafter\UrlBreaks\expandafter{\UrlBreaks\do\/\do\*\do\-\do\~\do\'\do\"\do\-}
\usepackage{upmath,color}
\usepackage{listings}
\usepackage{xcolor}

\lstdefinestyle{javastyle}{
    language=Java,
    basicstyle=\ttfamily\footnotesize,
    keywordstyle=\color{blue},
    commentstyle=\color{gray},
    stringstyle=\color{orange},
    numbers=none,
    breaklines=true,
    frame=single,
    showstringspaces=false,
    tabsize=4
}

\jvol{XX}
\jnum{XX}
\paper{8}
\jmonth{Month}
\jname{Publication Name}
\jtitle{Publication Title}
\pubyear{2021}

\begin{document}

\sptitle{Article Type: Description  (see below for more detail)}

\title{Keeping Models and Code in Sync: Roundtrip Engineering for Tactical Domain-Driven Design}

\author{Weixing Zhang}
\affil{Karlsruhe Institute of Technology, Karlsruhe, 76131, Germany}

\author{Mario Herb}
\affil{esentri AG, Karlsruhe, 76275, Germany}

\author{Wai Chung Dorothy Cheng}
\affil{esentri AG, Karlsruhe, 76275, Germany}

\author{Michael Wagner}
\affil{esentri AG, Karlsruhe, 76275, Germany}

\author{Bowen Jiang}
\affil{Karlsruhe Institute of Technology, Karlsruhe, 76131, Germany}

\author{Tianhai Liu}
\affil{Karlsruhe Institute of Technology, Karlsruhe, 76131, Germany}

\author{Anne Koziolek}
\affil{Karlsruhe Institute of Technology, Karlsruhe, 76131, Germany}

\markboth{THEME/FEATURE/DEPARTMENT}{THEME/FEATURE/DEPARTMENT}

\begin{abstract}
\looseness-1
Domain-Driven Design gives teams a shared vocabulary for complex business logic, but that vocabulary only stays useful as long as the model and the code agree with each other. In practice, they drift apart: code changes outpace the model, or model revisions never make it into the codebase. 
This paper presents JDomInO, a bidirectional synchronization toolchain for tactical DDD that keeps a Java codebase and its domain model connected through a shared metamodel, with the goal of keeping the two in sync as the system evolves. JDomInO generates Java code structure deterministically from a domain model (forward path) and reconstructs a domain model from existing Java code (reverse path). 
The forward path has been fully validated on a Hotel Management scenario covering all 12 building block types in the metamodel; the reverse path's mapping logic has passed unit testing, with end-to-end validation underway. We also outline how the structured domain model produced by JDomInO could serve as a precision context layer for AI code assistants, helping them respect aggregate boundaries and DDD semantics that raw source code alone does not convey.
\end{abstract}

\maketitle

\chapteri{D}omain-Driven Design (DDD) has become a widely adopted approach for building complex business software systems~\cite{evans2004domain}. At the tactical level, DDD maps business concepts directly to code structures through building blocks such as aggregates, entities, value objects, and domain events, enabling software models to reflect business reality. In practice, however, this alignment is difficult to sustain over time.

The root cause lies in the absence of automated synchronization between models and code. Development teams typically maintain a domain model in the early stages of a project, but as requirements evolve, developers modify code directly without updating the model. Conversely, when architects revise the model, the corresponding code may not be updated accordingly. Over time, models and code diverge, a phenomenon which we term \textit{model drift}. Model drift undermines the core value of DDD and makes codebases increasingly difficult to understand for developers joining the project.

Existing tools address this problem only partially. Unidirectional code generation tools support model-to-code generation but cannot synchronize changes made to the code back to the model~\cite{czarnecki2006feature}. Reverse engineering tools such as MoDisco can reconstruct structural information from code but do not understand DDD semantic boundaries~\cite{bruneliere2014modisco}. LLM-based code assistants such as GitHub Copilot, when used without structured architectural context, frequently produce code that violates DDD principles~\cite{barke2023grounded} due to their lack of awareness of aggregate boundaries or ubiquitous language.

This paper presents JDomInO, a bidirectional synchronization toolchain for tactical DDD. JDomInO connects the model layer and the code layer through an explicit domain metamodel, supporting deterministic Java code generation from domain models (forward path) and reverse reconstruction of domain models from existing Java code (reverse path). 
This same metamodel, we argue, does double duty: beyond keeping models and code aligned, it offers a structured, semantically rich layer that AI code assistants can use as architectural context, a direction we return to in Section ``Toward DDD-Aware AI Tooling''.
The forward path has been fully validated in a Hotel Management scenario comprising approximately 60 Java files and covering all 12 tactical DDD building block types defined in the metamodel. 
The mapping logic of the reverse path has been verified through independent unit tests, and end-to-end integration with the Hotel Management scenario is currently in progress. In addition, this paper discusses the potential of the JDomInO domain metamodel as a structured context layer for large language models, with the goal of improving the quality of AI-assisted DDD development by providing precise architectural semantics to code generation tools.

The remainder of this paper is structured as follows. Section 2 describes the model-code synchronization problem. Section 3 presents the architecture and implementation of JDomInO. Section 4 demonstrates the toolchain using the Hotel Management scenario. Section 5 discusses the vision of AI-aware DDD tooling. Section 6 addresses limitations and future work. Section 7 concludes the paper.

\section{The Model-Code Synchronization Problem}

Tactical DDD provides a precise language for software modeling. Aggregates define consistency boundaries; entities track business objects through unique identities; value objects encapsulate domain concepts in an immutable manner; domain events record changes in business state; and repositories provide persistent access to aggregates. Together, these building blocks serve as the code-level embodiment of the ubiquitous language, enabling business experts and developers to communicate based on a shared set of concepts.

However, code alone is often insufficient for grasping the overall domain structure of a complex system. When a system involves multiple bounded contexts, numerous aggregates, and their interrelationships, developers need an explicit domain model to communicate design decisions, align team understanding, and help new members build a mental map of the system quickly. As a result, many teams practicing DDD maintain a visual domain model alongside the code, expressing tactical building blocks and their relationships in a structured form.
Once an explicit domain model is introduced, keeping it synchronized with the code becomes a persistent engineering challenge. Existing tool solutions exhibit clear limitations in addressing this problem.

Unidirectional code generation tools take a domain model as input and produce an initial code skeleton. This approach offers some value at project inception, but the generation is a one-time act. Once developers modify the code, the tool has no awareness of those changes. The connection between model and code is permanently severed, and the responsibility for synchronization falls back to manual effort.

Reverse engineering tools can extract structural information from code, but such tools are typically designed for general object-oriented structures and lack the ability to recognize DDD semantics. They can recover dependencies between classes, but cannot determine which classes are aggregate roots, which are value objects, or where aggregate boundaries lie. Reconstructing a meaningful domain model from the output of these tools still requires substantial manual intervention.

Maintaining model-code consistency by hand relies entirely on team discipline. Under the pressures of rapid iteration, this approach is difficult to sustain and is easily disrupted by staff turnover, communication gaps, or delivery deadlines.

These limitations point to a common gap: no existing tool supports bidirectional synchronization between models and code while preserving an understanding of DDD semantics. This is the gap that the JDomInO toolchain aims to fill.

\section{JDomInO: A Roundtrip Engineering Toolchain}

\subsection{Architecture Overview}
JDomInO is a stage-wise engineering realization of the constraint-driven roundtrip engineering vision we proposed in prior work~\cite{zhang2026round}. This paper focuses on the core outcomes that have been realized: deterministic Java code generation from a DDD-native metamodel via the forward path, and reverse reconstruction of domain models from existing Java code via the reverse path. In contrast to the conceptual description in the vision paper, this paper reports the concrete implementation mechanisms of both paths, the scope of validated capabilities, and the technical limitations identified during implementation.

The central design idea of the toolchain is to use a DDD-native metamodel implemented in Java as a shared contract between the model layer and the code layer. Both forward generation and reverse reconstruction are mediated by this metamodel, which is designed to ensure semantic consistency across both directions. At the persistence layer, JDomInO separates semantic information from visual information across two storage files: \texttt{domain.json} records the semantic content of the domain model, including all tactical DDD building blocks and their relationships; \texttt{notation.json} records presentation information independent of semantics, such as the positions and layout of elements in the graphical interface. The motivation for this separation is that semantic information and presentation information have different lifecycles. Changes to code or model content affect the semantic layer; changes to layout or presentation affect the notation layer. Storing them together introduces unnecessary coupling and conflicts.

\begin{figure*}[tb]
  \centering
  \includegraphics[width=\linewidth]{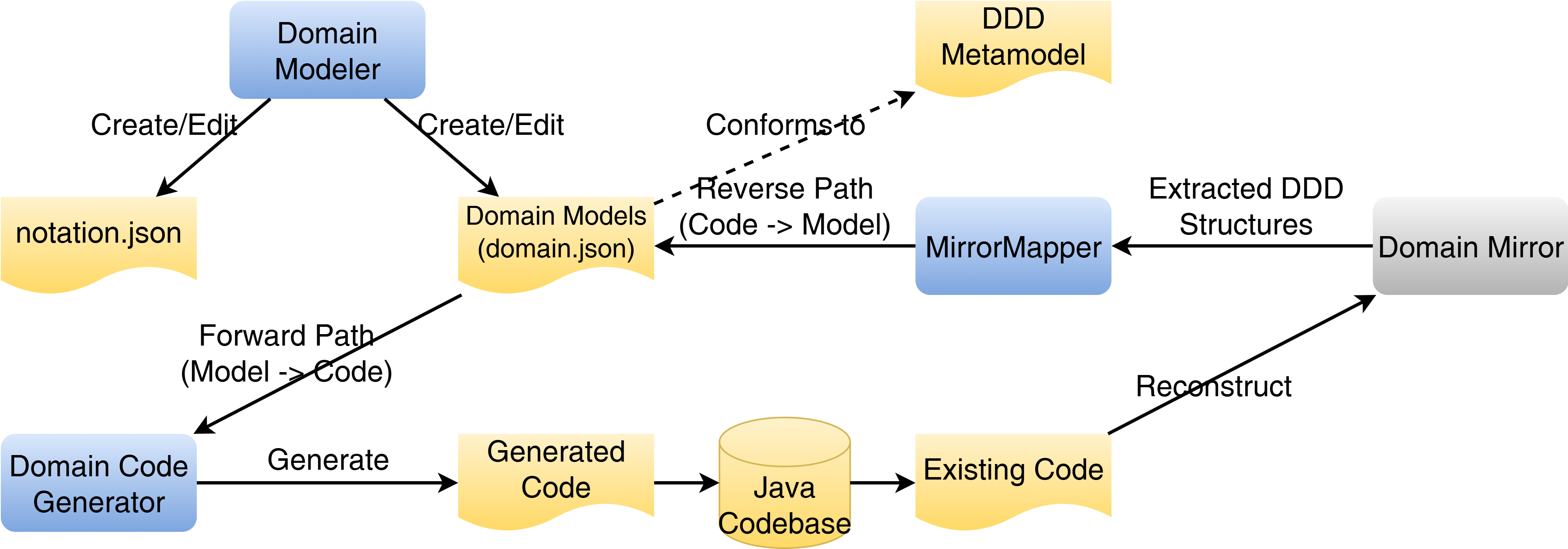}
  \caption{JDomInO Toolchain Architecture.}
  \label{fig:architecture}
\end{figure*}

The toolchain consists of two purpose-built components and one external dependency: the Domain Code Generator handles the forward path; MirrorMapper handles the second step of the reverse path; and Domain Mirror, an existing open-source component from esentri's DLC framework, handles the first step. Figure~\ref{fig:architecture} illustrates the overall architecture and data flow of the toolchain.

\subsection{Metamodel}
The JDomInO metamodel is defined as a set of Java classes and serves as the semantic foundation of the entire toolchain. In contrast to generic approaches that map DDD patterns to UML stereotypes~\cite{zhang2026round}, this metamodel treats tactical DDD patterns as first-class modeling primitives, enabling both code generation and model reconstruction to reason at the level of DDD semantics rather than generic object structures.

The metamodel covers 12 tactical DDD building blocks: AggregateRoot, Entity, ValueObject, DomainEvent, Repository, DomainService, ApplicationService, DomainCommand, Enum, QueryHandler, OutboundService, and ReadModel. The first six are the classical tactical DDD building blocks defined by Evans~\cite{evans2004domain}. The remaining six extend the metamodel toward modern architectural practices: DomainCommand, QueryHandler, and ReadModel correspond to the core roles of the Command Query Responsibility Segregation (CQRS) pattern; ApplicationService and OutboundService serve as the inbound and outbound ports of hexagonal architecture. DomainService supports more complete domain modeling as auxiliary types; Enum, treated as a distinct building block for code generation purposes, represents a design-time-bounded specialization of ValueObject.

The metamodel also explicitly encodes structural relationships between building blocks, including three representative categories: the compositional relationship within an aggregate (between AggregateRoot and the Entity and ValueObject instances it contains, annotated with multiplicity); the one-to-one architectural correspondence between a repository and its aggregate; and the publication relationship between a service method and the DomainEvent it publishes. At the semantic layer, these relationships are persisted as directed edges in \texttt{domain.json}. At the visual layer, they are expressed independently through node membership attributes. The two representations remain decoupled from each other, reflecting the consistency of the two-layer storage design.

\subsection{Forward Path: Domain Code Generator}
The forward path is implemented by the Domain Code Generator component, which reads a domain model from \texttt{domain.json} and generates the corresponding Java code deterministically. Determinism is the core design principle here: for the same model input, the generated code is always identical, independent of any external state. This property makes the output predictable, testable, and reproducible, and lays the foundation for incremental synchronization in future work.

The forward path has been fully validated in the Hotel Management scenario. The domain model for this scenario covers all 12 tactical DDD building block types defined in the metamodel, from which the Domain Code Generator produced approximately 60 Java files. All generated output passed character-by-character comparison against predefined golden files. This validation covers the complete set of building block types in the metamodel and provides credible empirical support for the correctness of the forward path.

\subsection{Reverse Path: Domain Mirror and MirrorMapper}
The reverse path aims to reconstruct a structured domain model from existing Java code. The process consists of two steps.

The first step is handled by Domain Mirror, an open-source component developed by esentri as part of their DLC framework~\footnote{\url{https://github.com/esentri/domainlifecycles/tree/main/mirror}}. Domain Mirror itself contains a metamodel of the tactical DDD structures within bounded contexts. It is initialized at application startup via Java reflection, mirroring the implementation state of DDD building blocks in the current codebase. The key value of Domain Mirror lies in its ability to recognize DDD semantics: it not only reconstructs the structural information of classes, but also identifies semantic roles such as aggregate roots, value objects, and domain events, providing a semantically rich intermediate representation for subsequent model reconstruction. It should be noted that for codebases making use of generics or deep inheritance hierarchies, Java's type erasure mechanism may prevent the reflection-based initialization from capturing complete type information, in which case additional configuration is required to ensure the completeness of type resolution.
In practice, however, the self-contained nature of a bounded context constrains generic usage to predictable structural patterns, allowing the toolchain to infer or defer nearly all generic type information within this boundary and effectively mitigating this limitation.

The second step is handled by the MirrorMapper component. MirrorMapper takes the intermediate representation produced by Domain Mirror and transforms it into the DomainModel structure defined by the JDomInO metamodel, which is then serialized to \texttt{domain.json}. The essence of this step is semantic adaptation: translating Domain Mirror's DDD-aware understanding into the standard representation of the JDomInO metamodel. The mapping logic of MirrorMapper has been verified through independent unit tests; in all tested scenarios, type names, fields, methods, and inter-building-block relationships were correctly reconstructed. End-to-end integration testing with the Hotel Management scenario is currently in progress.

The reverse path adopts a best-effort matching strategy for element identity rather than embedding persistent identity markers into source code; we discuss the resulting trade-offs in the Limitations section.

\section{Demonstration: Hotel Management System}
We demonstrate JDomInO using a hotel front-desk management system, a single bounded context built specifically to exercise the toolchain. The domain model comprises five aggregates (Hotel, Zimmer, Buchung, Rechnung, and ServiceLeistung), two entities, and seven value objects, alongside the CQRS- and hexagonal-architecture-oriented types the metamodel supports (DomainCommand, QueryHandler, ReadModel, ApplicationService, and OutboundService). This model exercises all 12 tactical DDD building block types defined in the metamodel, giving the forward path validation a broad footprint across the type system.

To illustrate the forward path concretely, we walk through Adresse, a value object attached to the Gast entity. In domain.json, Adresse is modeled with four string attributes: street, house number, postal code, and city. The Domain Code Generator translates this specification deterministically into the Java code shown in Listing~\ref{lst:adresse}.

\begin{lstlisting}[style=javastyle, caption={Generated code for the \texttt{Adresse} value object.}, label={lst:adresse}]
package com.esentri.rezeption.core.domain;

import lombok.Builder;
import jakarta.validation.constraints.NotBlank;
import io.domainlifecycles.domain.types.ValueObject;
import jakarta.validation.constraints.Size;

@Builder
public record Adresse(@NotBlank @Size(max = 200) String strasse, @NotBlank @Size(max = 20) String hausnummer, @NotBlank @Size(max = 10) String postleitzahl, @NotBlank @Size(max = 100) String ort) implements ValueObject {

    public Adresse {
    }
}
\end{lstlisting}

The generated class reflects several design decisions encoded in the metamodel at once: the ValueObject marker interface from the underlying DLC framework, a Lombok-generated builder, Jakarta Bean Validation constraints mirroring the attribute constraints in the model, and the immutability guaranteed by Java's record syntax, directly satisfying the semantic requirement that value objects remain immutable. Adresse is one of approximately 60 Java files generated for this scenario, spanning all 12 building block types from aggregate roots to application services; every generated file passed a character-by-character comparison against predefined golden files.

The model also captures relationships that are difficult to recover from code alone. For instance, when CheckIn, a domain service, completes a check-in operation, it triggers a BuchungEingecheckt event that ZimmerUseCases listens for to update room availability accordingly. Such event-driven dependencies are represented explicitly in domain.json, making them queryable at the model level 
without requiring repeated reflection-based analysis.

On the reverse path, MirrorMapper's mapping logic has been independently unit-tested against this scenario: type names, fields, methods, and inter-building-block relationships were all correctly reconstructed, giving us confidence in the correctness of the semantic adaptation step itself. What remains is end-to-end integration against the full Hotel Management codebase, which will validate the reverse path under the same conditions the forward path has already passed.



\section{Toward DDD-Aware AI Tooling}
\label{sec:ddd_ai}
The bidirectional synchronization architecture of JDomInO not only addresses the model-code consistency problem, but also lays the foundation for a broader research direction: using structured domain models as a precision context layer for AI code assistants. This section discusses the potential of this direction.

Current mainstream AI code assistants such as GitHub Copilot typically retrieve context from raw source code. In general-purpose programming scenarios, this approach has demonstrated considerable practical value. 
In DDD scenarios (i.e., scenarios with complex business requirements),
however, raw code context alone has a structural limitation: without additional architectural guidance, it cannot reliably convey architectural semantics such as aggregate boundaries, ubiquitous language, and tactical patterns to the AI~\cite{barke2023grounded}. When generating code, AI assistants can often reproduce the surface structure of the code, but without explicit architectural context, struggle to recognize which classes form an aggregate, which operations violate the immutability of a value object, or which service should publish a given domain event. Existing research indicates that this absence of architectural semantic awareness is one of the contributing factors to the persistent divergence between models and code in DDD practice~\cite{ozkan2025domain}. Moreover, even when architectural rules are injected via structured context files, AI agents do not always strictly adhere to them, producing outputs that may violate DDD constraints in unpredictable ways. 
This is a difference in kind, not degree: where such context files guide AI agents through natural-language description of intent, JDomInO's metamodel encodes the same rules as deterministic structure, leaving no room for probabilistic interpretation.
A deterministic validator component that checks AI-generated code against the domain model, as envisioned in future work, could provide reliable post-generation checks and deliver direct feedback to both human developers and AI agents, closing this loop.

The \texttt{domain.json} artifact produced and maintained by JDomInO has unique potential in this context. Compared to raw Java code, \texttt{domain.json} encodes complete domain semantics in a compact, structured form: the types, attributes, and relationships of all building blocks, as well as aggregate boundaries, are explicitly represented. Certain semantic relationships are particularly difficult to recover from source code alone; domain events, for instance, are typically injected into the control flow rather than invoked as direct method calls, making their publication relationships hard to trace through static dependency analysis.
This means that the same semantic information can be conveyed to an LLM with fewer tokens, at lower noise, and with greater structural clarity. 
This aligns with a recent large-scale mapping study of LLM applications in model-driven engineering, which found that lightweight textual formats are more often preserved across LLM-based workflows than native MDE exchange formats such as XMI/Ecore~\cite{zhang2027large}.
Furthermore, the structural boilerplate generated deterministically by the forward path requires no LLM involvement, reducing token consumption for the code generation task itself.
More importantly, the naming used in \texttt{domain.json} derives directly from the ubiquitous language, naturally aligning with the terminology used in business requirement descriptions. This provides a more reliable semantic bridge for LLMs mapping business requirements to correct technical implementations.

This observation gives rise to several research directions worth exploring. First, does using \texttt{domain.json} as the primary context for an LLM, in place of raw source code, reduce architectural semantic errors in AI-generated code while also lowering token consumption? Second, by feeding AI-generated code back through JDomInO's reverse path and reconstructing the domain model, is it possible to conduct a high-level consistency audit of the AI's architectural contributions at the model layer, detecting violations of DDD constraints more efficiently than traditional code review? Third, does a hybrid workflow that combines deterministic code generation for structural DDD boilerplate with AI generation for complex business logic outperform pure AI-based generation in terms of architectural stability and development efficiency?

The infrastructure established by JDomInO provides unique research conditions for exploring these directions: the forward path ensures deterministic generation of structural code; the reverse path provides a mechanism for mapping AI-generated code back to the model layer; and the unified domain metamodel serves as a semantic anchor for AI context engineering.


\section{Limitations and Future Work}

The current implementation of JDomInO has several known boundaries. On the reverse path, the toolchain adopts a best-effort element-matching strategy rather than embedding persistent identity markers in source code; as a result, class renaming or relocation 
still requires manual reconciliation, though for different reasons: a deliberate trade-off on the code side to avoid embedding toolchain artifacts, and an open implementation gap on the model side where orphaned files are not yet cleaned up.
In addition, domain.json currently lacks a complete independent model-integrity validation step: consistency relies primarily on the correctness of the generation and reconstruction pipelines themselves, though a set of consistency APIs is under development to enforce model correctness at the operation level, preventing structurally invalid configurations from being introduced through the modeler.
Finally, the reverse path's end-to-end validation against the Hotel Management scenario is the next milestone on our roadmap: unit-level correctness has been established, and full-scenario integration is scheduled to close this gap in the coming months.

Building on these observations, we plan to pursue two directions going forward. First, we intend to introduce a deterministic validator component that actively checks the integrity of domain.json and provides automated architectural-compliance feedback for AI-generated code, a capability that is also a prerequisite for the AI-aware closed loop envisioned in the previous section. Second, we plan to pursue the research questions raised around domain.json as a structured context layer for LLMs, including its effect on token efficiency and architectural-semantic accuracy, and whether a hybrid workflow combining deterministic generation with AI-based generation outperforms pure AI-based generation; this remains the direction we consider most promising for long-term impact. In parallel, we are conducting a large-scale empirical study of DDD practice in open-source repositories, investigating how Bounded Context boundaries erode over time~\cite{zhang2026domain}; its findings are expected to directly inform the constraint rules underlying JDomInO's future validator component.

\section{Conclusion}

Model drift is not a failure of discipline; it is a failure of tooling. JDomInO is designed so that a shared, DDD-native metamodel can keep code and model in sync in both directions, not just at project inception but as an ongoing property of the codebase. The forward path has been fully validated; the reverse path is close behind. What we see on the horizon is just as important as what we have built: a deterministic validator that can hold AI-generated code to the same architectural standards as human-written code, and a domain model that speaks the language of the business well enough to guide, rather than just describe, that generation. Every team practicing DDD at scale eventually faces model drift. We believe this should be treated as a tooling problem, not a documentation problem, and JDomInO is our attempt to make that shift real. We invite practitioners to try it and help shape where it goes next.











\section{ACKNOWLEDGMENTS}
This work was funded by the German Federal Ministry of Research, Technology and Space (BMFTR) under grant no. 01IS25017B as part of the JDomInO project.


\def\refname{REFERENCES}

\bibliographystyle{IEEEtran}
\bibliography{bibliography}

\begin{IEEEbiography}{Weixing Zhang}{\,} is a postdoctoral researcher in the MCSE group at the Karlsruhe Institute of Technology, Germany. He received his Ph.D. degree from the University of Gothenburg in October of 2025. Before that, he worked in industry for 7 years as a software engineer, and he has extensive software development skills and experience. His research interests include empirical software engineering, MDE and DSL, and AI4SE.  \vspace*{8pt} 
\end{IEEEbiography}

\begin{IEEEbiography}{Mario Herb}{\,} is a co-founder of esentri AG in Karlsruhe, Germany, and he is a senior software developer and architect by heart. He is a Domain-Driven Design enthusiast and the project manager and lead designer of the project JDomInO. Contact him at 
mario.herb@esentri.com.\vspace*{8pt}
\end{IEEEbiography}

\begin{IEEEbiography}{Wai Chung Dorothy Cheng}{\,} is a software consultant at esentri AG (part of the Gofore Group) in Karlsruhe, Germany with 2 years of industry experience. She has a strong interest in software architecture and is an architect and a developer in the JdomInO project. Contact her at dorothy.cheng@esentri.com.\vspace*{8pt}
\end{IEEEbiography}

\begin{IEEEbiography}{Michael Wagner}{\,} is a Cloud Platform Consultant at esentri AG and holds an M.Sc. in Computer Science from the Karlsruhe Institute of Technology (KIT), Germany. In the JDomInO project, he is responsible for designing and implementing GLSP-based modeling tools to support automated round-trip engineering and domain model consistency. Contact him at michael.wagner@esentri.com.\vspace*{8pt}
\end{IEEEbiography}

\begin{IEEEbiography}{Bowen Jiang}{\,} is a Ph.D. researcher in the MCSE group at the Institute of Information Security and Dependability (KASTEL) at Karlsruhe Institute of Technology (Karlsruhe, Germany) since October 2024. Before that, she received the M.Sc. in Computer Science (Fundamental Science and Engineering) from Waseda University (Tokyo, Japan) in 2024. Her research interests include model-driven engineering, software testing, software maintenance, and AI4SE. Contact her at bowen.jiang@kit.edu.\vspace*{8pt}
\end{IEEEbiography}

\begin{IEEEbiography}{Tianhai Liu}{\,} is a postdoctoral researcher at Karlsruhe Institute of Technology. His research focuses on formal methods, software verification, model-driven engineering, and consistency management for cyber-physical systems. He has over ten years of industry experience in safety-critical systems and has led R\&D activities in European collaborative projects. Contact him at tianhai.liu@kit.edu.\vspace*{8pt}
\end{IEEEbiography}

\begin{IEEEbiography}{Anne Koziolek} {\,}  is a professor at Karlsruhe Institute of Technology. She is the head of the research group Modeling for Continuous Software 
Engineering (MCSE) at KASTEL – Institute of Information Security and Dependability. She is interested in conciliating model-based software 
engineering with development processes that have fast and agile feedback cycles and thus combine the benefits of both approaches. In particular, she is interested in tool support for systematic, yet low-cost model-based design space exploration to support making good design decisions.

\end{IEEEbiography}

\end{document}